\documentclass[twocolumn]{aastex63} 

\newcommand{\lasp}{
	Laboratory for Atmospheric and Space Physics,
	University of Colorado at Boulder,
	3665 Discovery Dr.,
	Boulder, CO 80303-7814, USA
	}

 \newcommand{\cu}{ 
	University of Colorado at Boulder,
	Boulder, CO 80309-0390, USA
	}
\newcommand{\jila}{
	JILA, National Institute of Standards and Technology and University of Colorado, 
	Boulder, Colorado 80309, USA
	}
\newcommand{\jhuapl}{
	Applied Physics Laboratory,
	Johns Hopkins University,
	11100 Johns Hopkins Rd.,
	Laurel, MD 20723, USA
	}
\newcommand{\gsfc}{
	NASA Goddard Space Flight Center,
	8800 Greenbelt Rd.,
	Greenbelt, MD 20771, USA
	}
 \newcommand{\cires}{
	Cooperative Institute for Research in Environmental Sciences,
	University of Colorado,
	Boulder, CO 80309, USA
	}
	
\newcommand{\ncei}{
	NOAA National Centers for Environmental Information,
	325 Broadway,
	Boulder, CO 80305, USA
	}

\received{2023 March 3}
\revised{2023 April 10}
\accepted{2023 April 11}

\submitjournal{ApJ}

\shorttitle{Solar Flare Frequency Distribution}
\shortauthors{Mason et al.}

\graphicspath{{./}{figures/}}

\begin{document}

\title{Coronal Heating as Determined by the Solar Flare Frequency Distribution Obtained by Aggregating Case Studies}

\correspondingauthor{James Paul Mason}
\email{james.mason@jhuapl.edu}

\author[0000-0002-3783-5509]{James Paul Mason}
\affiliation{\lasp}
\affiliation{\jhuapl}

\author[0000-0003-0310-2654]{Alexandra Werth}
\affiliation{\cu}
\affiliation{\jila}

\author[0000-0003-4181-0889]{Colin G. West}
\affiliation{\cu}

\author[0000-0002-1176-3391]{Allison A. Youngblood}
\affiliation{\lasp}
\affiliation{\gsfc}

\author[0000-0001-9351-733X]{Donald L. Woodraska}
\affiliation{\lasp}

\author[0000-0002-7586-4220]{Courtney L. Peck}
\affiliation{\cires}
\affiliation{\ncei}
\affiliation{\lasp}

\author[0000-0002-8995-2408]{Arvind J. Aradhya}
\affiliation{\cu}

\author{Yijian Cai}
\affiliation{\cu}

\author[0000-0002-2628-7439]{David Chaparro}
\affiliation{\cu}

\author{James W. Erikson}
\affiliation{\cu}

% ...

\collaboration{6}{(Physics 1140 Class and TAs)} 

\author[0000-0002-0995-552X]{H. J. Lewandowski}
\affiliation{\cu}
\affiliation{\jila}

\begin{abstract} 
Flare frequency distributions represent a key approach to addressing one of the largest problems in solar and stellar physics: determining the mechanism that counter-intuitively heats coronae to temperatures that are orders of magnitude hotter than the corresponding photospheres. It is widely accepted that the magnetic field is responsible for the heating, but there are two competing mechanisms that could explain it: nanoflares or Alfv\'en waves. To date, neither can be directly observed. Nanoflares are, by definition, extremely small, but their aggregate energy release could represent a substantial heating mechanism, presuming they are sufficiently abundant. One way to test this presumption is via the flare frequency distribution, which describes how often flares of various energies occur. If the slope of the power law fitting the flare frequency distribution is above a critical threshold, $\alpha=2$ as established in prior literature, then there should be a sufficient abundance of nanoflares to explain coronal heating. We performed $>$600 case studies of solar flares, made possible by an unprecedented number of data analysts via three semesters of an undergraduate physics laboratory course. This allowed us to include two crucial, but nontrivial, analysis methods: pre-flare baseline subtraction and computation of the flare energy, which requires determining flare start and stop times. We aggregated the results of these analyses into a statistical study to determine that $\alpha = 1.63 \pm 0.03$. This is below the critical threshold, suggesting that Alfv\'en waves are an important driver of coronal heating.
\end{abstract}


\section{Introduction} 
\label{sec:intro}

It has long been established that the slope of a power law fitting the solar flare frequency distribution (occurrence rate versus energy) is a strong indicator of whether nanoflares are an important coronal heating mechanism (e.g., \citealt{Hudson1991, Crosby1993, Veronig2002}). The conceptual reasoning is simple: a large slope means that there is a large abundance of the smallest flares. These small flares may not individually transport much energy, but, in aggregate, they represent a substantial heat transfer mechanism. It is well established in the flare literature that many flare parameterizations tend to have power law relationships with each other (e.g., \citealt{Kahler1982, Veronig2002, Aschwanden2012, DHuys2016}). The flare frequency distribution (FFD) is one one such relationship well described by a power law: 

\begin{equation}
	\label{eq:alpha}
	\frac{dn}{dE} = AE^{-\alpha}
\end{equation}

\noindent where $n$ is the number of events, $E$ is the radiated flare energy, $A$ is an offset constant, and the primary parameter of interest is the exponent, $\alpha$, which manifests as a slope when fitting a line in log-log space. If $\alpha < 2$, then nanoflares are not frequent enough to contribute a sufficient amount of energy to the corona to explain its observed ambient temperature \citep{Hudson1991}. Thus, the accurate determination of $\alpha$ is of critical importance for addressing the longstanding coronal heating problem. 

Moreover, important points of comparison can be made between solar and stellar FFDs that can lead to greater insights into the mechanisms that drive the FFD and dominate heating mechanisms (e.g., \citealt{Loyd2018}). For example, many stellar studies find $\alpha < 2$ -- consistent with much of the solar FFD literature (e.g., \citealt{Crosby1993} and 11 references in their Table 1) -- and conclude that the underlying mechanism driving flaring and coronal heating is therefore the same as that of the Sun (e.g., \citealt{Airapetian2019} and references therein). Still other studies find stellar FFDs with $\alpha > 2$ (e.g., \citealt{Maehara2012}) -- also consistent with some of the solar literature (e.g., \citealt{Veronig2002}). We also know that flares of all energies (including superflares with $>$10$^{34}$ erg) tend to occur more often for rapidly rotating stars \citep{Maehara2012, Shibayama2013}.

There is a large body of research that has focused on the determination of $\alpha$ for the Sun. The ultimate goal is to obtain the total amount of radiated energy from each flare, excluding background radiated energy. This is a challenge. Most studies, including this one, focus on a single wavelength regime. Soft X-rays (SXRs) are a popular choice (e.g., \citealt{Hudson1969, Drake1971, Veronig2002, Aschwanden2012}) because during a flare they experience the greatest enhancement above their background levels and are therefore easier to detect and characterize \citep{Woods2004, Rodgers2006}, and they may be a better indicator of total flare energy release than, e.g., hard X-rays (HXRs; \citealt{Veronig2002, Lee1993, Feldman1997}). Furthermore, there are now several decades of continuous measurement of the SXR wavelength region from the NOAA Geostationary Operational Environmental Satellites (GOES) X-ray Sensor (XRS) in two channels; XRS-A from 0.5-4~\AA (AKA ``short") and XRS-B from 1-8~\AA (AKA ``long"). \citet{Crosby1993} list many studies and which wavelength regime each focused on to determine $\alpha$ in their Table 1. Some studies have analyzed the global energetics spanning multiple wavelengths and accelerated particles, but this detailed analysis typically can only be done for a relatively small number of events, for example, the 38 large events studied by \citet{Emslie2012}. Some studies calculate the radiated energy in their selected wavelength regime (e.g., \citep[e.g.,][]{Shimizu1995}, while others do not, instead focusing on peak irradiance or other parameters to act as a proxy for the energy (e.g., \citealt{Aschwanden2012}). Furthermore, some studies have subtracted the background irradiance from the flare, while others have not. \citet{Veronig2002} found that this could have an impact on the derived value of $\alpha$, in that case bringing it from $\alpha = 2.03 \pm 0.09$ without subtraction to $\alpha = 1.88 \pm 0.11$ with subtraction -- straddling the critical value of $\alpha = 2$. 

While it would be ideal if these studies were done spanning all wavelengths, that is an extremely challenging task that would not necessarily lead to different conclusions given that SXRs have already been established to be a reliable indicator of the total energy release, which is partially why the GOES/XRS-B peak measurement remains the de facto definition for flare magnitude. In summary, it is important to study a large number of flares, subtract the background irradiance, and calculate the energy. That is precisely what has been done for the present study. 

A novel aspect of this study is the large number of contributing authors, which we call the Colorado Physics Laboratory Academic Research Effort (C-PhLARE) Collaboration. Over 1400 undergraduate students participated in an introductory physics lab course run as a Course-based Undergraduate Research Experience (CURE) at the University of Colorado Boulder \citep{Corwin2014}. The students were enrolled in the CURE in one of three semesters from Fall 2020 through Spring 2021. The students worked in small teams on this research, where they chose individual flares, removed the background irradiance, and computed the energy of single flares. Their analysis was peer-reviewed by other teams in the class and iteratively improved based on this feedback. The students then submitted their final results in memo form as a flare archive entry, which included the calculated total energy of the flare, the peak irradiance, beginning and ending times of the flare, and plots of the analysis. These memos were used by the senior authors to review the analysis and determine which flares analyses met the standards to be included in the final dataset used for the work presented here. This process is described in detail in Section \ref{sec:datareview}. A more detailed description of the course, educational objectives, and results of student learning can be found in \citet{Werth2022a}, \citet{Werth2022b}, and \citet{Werth2022c}.

All students who were enrolled in the class were offered authorship. To be included in the final author list, students had to have completed the course during one of three semesters this project was run, earned a passing grade in the class, and opted-in to being an author through a web-form. This process resulted in 964 student co-authors. Additionally, all graduate teaching assistants for the course during these semesters were offered authorship, resulting in 31 additional co-authors. 

The paper begins by describing the data used for this study (Section \ref{sec:data}), which is predominately from GOES/XRS with a calibration check provided by the Miniature X-ray Solar Spectrometer (MinXSS; \citealt{MasonMinXSS2016}). Section \ref{sec:methods} describes the methods applied to study the flares in our dataset, which include background subtraction and energy calculation. Section \ref{sec:results} presents the resultant FFD and comparisons with other solar and stellar work. Section \ref{sec:discussion} closes with a discussion. 

\section{Data}
\label{sec:data}

\begin{deluxetable}{ccc}
	\tablecaption{Key instrument/dataset characteristics}
	\label{tab:instruments}
	\tablehead{\colhead{} & \colhead{GOES/XRS} & \colhead{MinXSS-1}} 
	\startdata
		Bandpass (native) & \begin{tabular}[c]{@{}l@{}} XRS-A: 0.5-4 \AA (3-25 keV)\\ XRS-B: 1-8 \AA (1.5-12 keV)\end{tabular} & 0.4-30 \AA (0.4-30 keV) \\
		Bandpass used & 1-8 \AA (1.5-12 keV) & 0.4-30 \AA (0.4-30 keV) \\
		Spectral resolution & -- & 0.15 keV \\
		Cadence (native) & 2 s & 10 s \\
		Cadence used & 1 minute & 1 minute\\
		Dates used & \begin{tabular}[c]{@{}l@{}}2011-01-14 to \\ 2018-02-10\end{tabular} & \begin{tabular}[c]{@{}l@{}}2016-05-16 to \\ 2017-05-06\end{tabular} \\
            Data product & xrsf-l2-avg1m\_science & Level 1 \\
            Data version & v1-0-0 & v4.0.0
	\enddata
\end{deluxetable}

Two primary sources of data were used for this study: measurements from GOES-15/XRS and MinXSS-1 (see Table \ref{tab:instruments}). Prior to GOES-16, the XRS instruments did not have an absolute calibration. MinXSS-1, however, flew concurrently and observed the same bandpass as GOES/XRS (and beyond), but with spectral resolution and an absolute calibration \citep{Moore2016} obtained at the National Institute of Standards and Technology (NIST) Synchrotron Ultraviolet Radiation Facility (SURF), so we used its data as a validation check of the GOES-based results. MinXSS-1 data are limited to a single year (2016 May - 2017 May) and the spacecraft experienced regular ($\sim$15/day) eclipses that caused gaps in some flare observations. GOES/XRS has provided nearly continuous X-ray observations since 1975 from multiple generations in the satellite series with minimal measurement interruptions from eclipses due to a geostationary orbit. We focused on GOES-15 because 1) it spans the entirety of the most recent solar cycle, 2) it overlaps with the MinXSS-1 observations, and 3) focusing on a single satellite in the series avoids any concerns with cross-calibration. This study uses the science-quality GOES-15 XRS data products from NOAA's National Centers for Environmental Information (NCEI)\footnote{ \url{https://www.ncei.noaa.gov/data/goes-space-environment-monitor/access/science/xrs/GOES_1-15_XRS_Science-Quality_Data_Readme.pdf}} available as of early 2020. Compared to the previously available GOES-15 XRS data, the science-quality data uses a significantly improved background correction and remove scaling factors applied by the Space Weather Prediction Center (SWPC) to the XRS channel irradiance measurements (0.85 and 0.7 for XRS-A and -B, respectively) to provide the irradiance measurements in true physical units of W m$^{-2}$.

\begin{figure}
\plotone{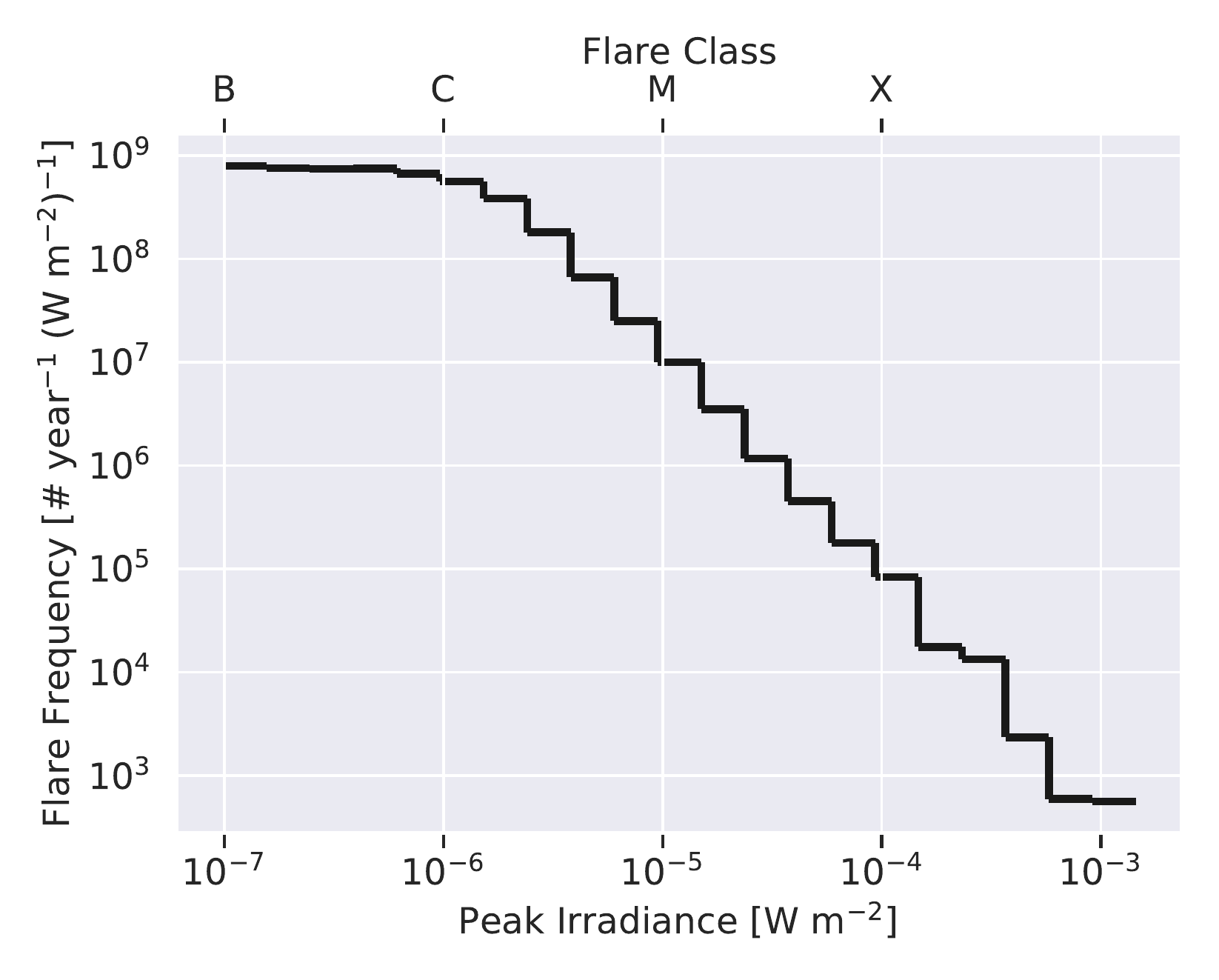} 
\caption{Histogram of the 18,833 flares identified by NCEI for the time range specified in Table \ref{tab:instruments} using logarithmic bin widths. Note that the saturation at the smallest flare classes is due to low signal-to-noise ratio in the instrument, but is also due to the Sun's inherent background intensity, which can obscure the smallest classes of flares. The slope of the power law represented here is $\alpha_{peak\_irrad} = 2.16 \pm 0.01$.
\label{fig:flare_class_hist}}
\end{figure}

GOES/XRS L2 1-minute average data were retrieved from the Space Weather Data Portal (SWDP; \citealt{swxdataportal, Knuth2020}) and data for MinXSS are available from NASA's Solar Data Analysis Center\footnote{\url{https://sdac.virtualsolar.org/cgi/show_details?provider=LASP_MINXSS1}}. A total of 18,833 flares were identified in the 1-minute average GOES/XRS-B irradiances by the NCEI L2 flare summary algorithm\footnote{The GOES-15 XRS L2 data products are generated using the same algorithms as those used for the next-generation GOES-R (GOES-16 through -19) XRS data products. Details of the L2 algorithms can be found at \url{https://data.ngdc.noaa.gov/platforms/solar-space-observing-satellites/goes/goes16/l2/docs/GOES-R_XRS_L2_Data_Users_Guide.pdf}} in the time period listed in Table \ref{tab:instruments}; Figure \ref{fig:flare_class_hist} shows the corresponding distribution of flare classification. This flare classification nomenclature is an alphanumeric labeling of the flare's 1-minute-averaged peak flux measured by GOES/XRS-B. To determine the slope of the power law represented in Figure \ref{fig:flare_class_hist}, we followed the the recommendation in \citet{DHuys2016} to use a Maximum Likelihood Estimator (MLE). They found that MLEs converge much faster as a function of sample size than other methods. This was not crucial for the data in Figure \ref{fig:flare_class_hist} given the large number of events (a linear regression results in a similar value of $\alpha_{peak\_irrad} = 2.22 \pm 0.07$), but the MLE method will be important when we determine the slope of the energy FFD ($\alpha$) in Section \ref{sec:ffd_method}, where the number of events is on the order of $\sim$200.

\section{Methods} 
\label{sec:methods}

The overarching procedure applied to the GOES/XRS data are outlined here and detailed in subsequent subsections. 

Working in teams of three or four, students on the project completed the following steps: 

\begin{enumerate}
	\item Selected a flare to analyze, under the constraint that we needed to ensure broad coverage across the entire solar cycle and across flare classifications
	\item Subtracted pre-flare baseline level
	\item Integrated light curve over duration of flare
	\item Converted to energy units
	\item Cross validated another team's result
	\item Combined the results to produce a dataset of the total energy and peak irradiance of each flare
\end{enumerate}

After the above steps were completed by the student research teams, senior researchers on the project completed an individualized review of these teams' analyses and the resulting data, addressing items such as duplicate flares or faulty calculations. 

Finally, the resulting dataset was used to produce a flare frequency distribution and determine the best-fit value of $\alpha$.

Details for each of these steps are outlined in subsections below. We also identified three flares that received this treatment and were also observed by MinXSS-1, which we analyzed in order to ``cross check" the absolute energy calibration. This MinXSS analysis method is further detailed in Section \ref{sec:minxss_method}. 

\subsection{Flare Selection}

Flares were selected for analysis by teams of 3-4 students over three semesters, with instruction for some teams to select different class flares to ensure we sampled the full distribution. During the first semester, students were permitted to choose from flares across the entire lifetime of the GOES-15 satellite, 2010 March through 2018 December. The students chose flares mostly close to the solar minimum. In subsequent semesters, students were asked to focus on flares closer to the solar maximum to make sure there was coverage of the entire solar cycle. The resulting final dataset contains flares with a majority near the solar maximum -- though we note that prior studies have searched for and failed to find any significant solar cycle dependence in the FFD slopes (e.g., \citealt{Veronig2002, Aschwanden2012}).

In addition, there were other factors that guided student flare selection. For example, students were instructed to choose single flares that did not overlap with other flares of similar or greater irradiance. Students may have looked at the pre-flare background and chosen flares with a flatter background for simpler pre-flare baseline subtraction. There were cases where multiple student teams analyzed the same flare. If a flare had been analyzed more than twice, the research leads selected the one closest to the mean of the set; while the flares that were analyzed exactly twice were down-selected randomly. 

\subsection{Pre-flare Baseline Subtraction}

As described in Section \ref{sec:intro}, pre-flare baseline subtraction is a crucial step because it reduces the impact of the baseline level of X-rays being counted as part of the energy of the flare, which has been shown to cause the calculation of $\alpha$ to straddle the critical value of 2. While the concept of pre-flare baseline determination is simple, in practice, it is not trivial. It is common that there are either other flares just prior to the flare-of-interest or minutes-to-hours-long trends in the irradiance due to active region evolution. Automated methods can be developed for this determination, but tend to rely on algorithms with fixed thresholds for, e.g., how much variation is allowed in the pre-flare baseline. As a result, many flares must be rejected in the automated analysis for violating these thresholds, which we refer to as ``low throughput" for the algorithm. In \citet{Mason2019}, this throughput was 30\% for pre-flare baseline determination; subsequent steps in the automated algorithm had rejection criteria of their own that further reduced the throughput of the full algorithm. Manual determination of the pre-flare baseline level can have a higher throughput (see Section \ref{sec:datareview}), but tends to limit the number of events that can be studied and, thus, the statistical significance of the final result. Our solution is to distribute the task across a large number of undergraduate researchers. 

\begin{figure}
\plotone{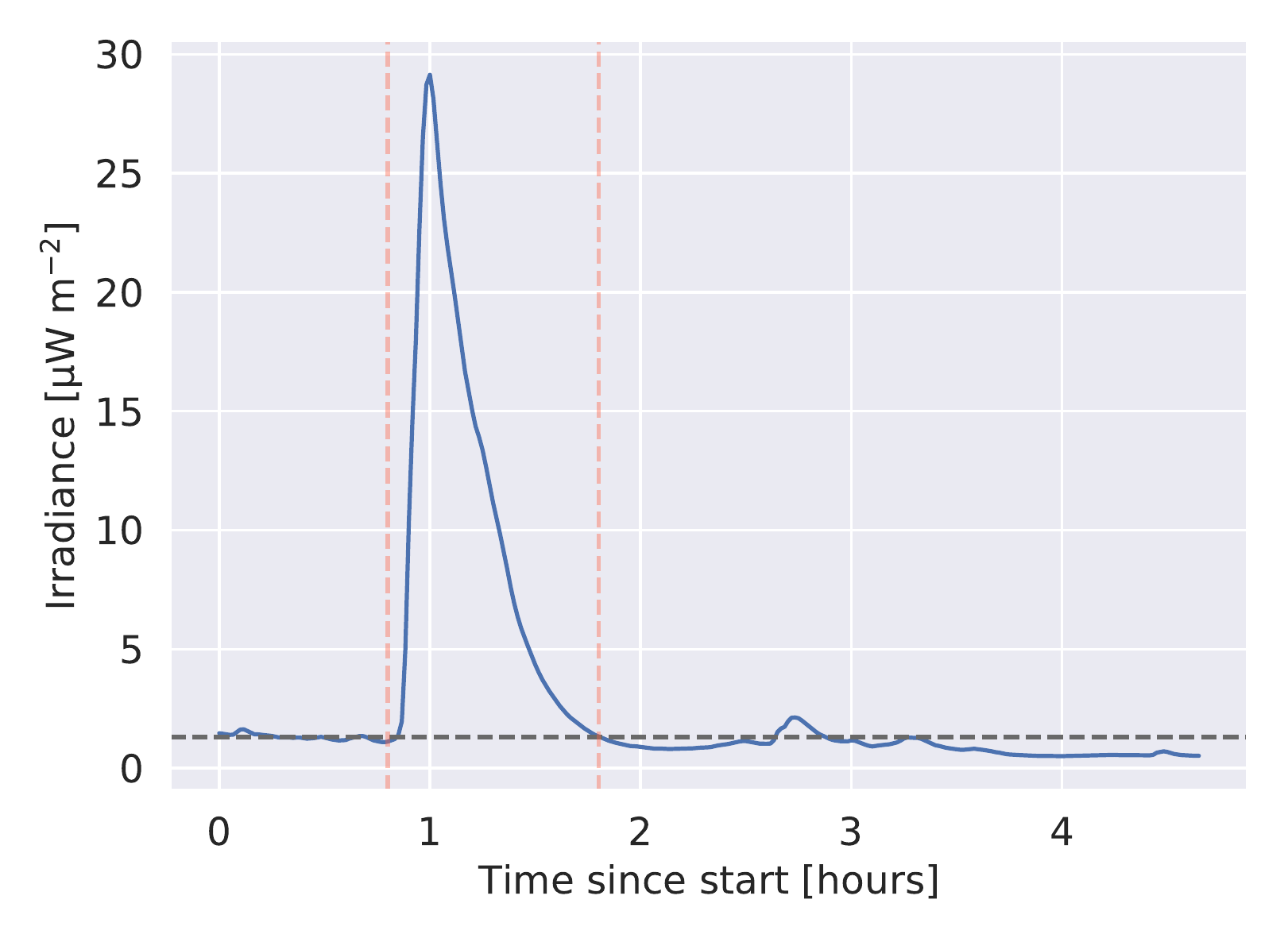} 
\caption{Example GOES/XRS light curve with annotations for the baseline level (horizontal dashed) and start/stop times (vertical dashed) as determined by the student team for this particular event. This particular flare peaked as an M2 at 2016-07-24 06:20:00. Simple light curves like this one are ideal, but they can become much more complex (e.g., see \citealt{Mason2019}).
\label{fig:example_light_curve}}
\end{figure}

To determine the pre-flare background, teams selected from a variety of methods depending on the features of the particular flare in question. Most were able to identify a period shortly before the start of the flare that did not contain additional small flare signals, and was reasonably long compared to the timescale of the target flare (e.g., Figure \ref{fig:example_light_curve}). When this was not possible, teams chose a method to exclude small flare signals before averaging -- for example, clipping the data at the level of a B-class flare, or averaging only over the local minima in a pre-flare region. 

\subsection{Flare Duration Integration}

Integrating the light curve over time is another step that is simple in concept, but non-trivial in practice. The start and stop times have to be chosen. Additionally, flare profiles are not always ``clean": due to subsequent flares or active region evolution and the profile may not return directly to the pre-flare baseline. The NCEI L2 flare summary algorithm flags a flare end time for every event: whatever time the irradiance drops to half the peak value. This is a consistent but problematic metric for our application because it ignores valid energy that was released by the flare, though the end time is not as important as the start time for determining the total energy \citep{Ryan2016}. Just as with pre-flare baseline determination, this step can also be automated, but results in major caveats in the conclusion; yet manual determination for every flare is cumbersome. Again, the advantage of the present study is the large number of researchers that can distribute the task. As with the background subtraction, teams made individualized determinations about the start and end points of the integration. As a guide, many began by examining points where the flare crossed the previously-determined background level, but often made small adjustments based on the nuances in the data for that particular event. Note that the units of the GOES/XRS data are W m$^{-2}$ = J s$^{-1}$ m$^{-2}$. This time-integration step eliminates the s$^{-1}$, leaving us with J m$^{-2}$. 

\subsection{Conversion to Energy Units}

In the GOES/XRS irradiance units, the m$^{-2}$ is a normalization of the flux to a shell with radius equal to the Sun-Earth distance at the time of observation, which is 1 au on average. To get energy independent of any particular distance scaling, we simply multiply by 4$\pi$d$^2$, where d is the Sun-Earth distance at the time of observation, then convert to ergs for easier comparisons with prior studies. 

\subsection{Cross Validation}

After each team had completed at least one flare analysis, teams exchanged analysis documents to provide dual anonymous reviews of each other's work. Each flare analysis received reviews from multiple other students. Reviewers identified any clearly erroneous work (for example, mistake in units conversion or a coding error) and assessed the reasonableness of the methods used for the baseline analysis and integration. In many cases, a variety of methods would be reasonable and produced very similar numbers, but the review process helped, in part, to identify cases where the choice of method was dramatically impacting the results. The teams then received the feedback and made improvements to their analysis before reporting the results to the collaboration. 

Finally, for each flare, the student teams created an entry in a ``flare archive," which was a formal presentation of their analysis documents and included basic information about the event (e.g., timestamps, classification, reference solar image), a description of the flare's environment (e.g., solar minimum, close in time to other flares or not), and a description of the specific methods applied in Steps 2 and 3, with accompanying plots. 

\subsection{Data Review}
\label{sec:datareview}

After the student data was compiled in our ``flare archive," the senior researchers on the project were able to review the work flare-by-flare as an additional check above and beyond the review process that took place among the student teams. This comprehensive data review involved the following steps: 

\begin{itemize}
    \item Resolving instances of duplicate flares, where multiple student teams chose the same flare to analyze.
    \item Ensuring that the peak irradiance and other details reported for each flare matched the data in the SWDP.
    \item Removing flares from the database that contained an overlapping flare.
    \item Removing flares from the database where the pre-flare background displayed a secular trend that would make the background subtraction unreliable.
\end{itemize}

The flare archive included 607 entries; removal of duplicates left us with 350 data points; removal of unreliable data from the remaining checks left 212, for an overall throughput of 35\%. We note that this compares favorably with attempts to use automated algorithms to analyze similar light curves: in a recent example, throughput for pre-flare baseline identification and event start/stop times was only 17\% \citep{Mason2019}.

\subsection{Flare Frequency and Determination of $\alpha$}
\label{sec:ffd_method}

Essentially all studies of flare frequency distributions consider some sub-sample of all flares that occurred. Depending on the selection method, this can inject bias. The present study also had to sub-sample, and, here, we describe how we corrected the bias. Because the flare analysis used a stratified flare selection process to ensure adequate sampling (i.e., some teams of student researchers were instructed to choose an X-class flare, others an M-class, etc), the frequency of various flare types in our dataset is unnaturally skewed. We cannot rely, therefore, on the frequency of various flare types within our sample as a reliable indicator of the actual frequency of such flares in nature. To correct for this when producing our final FFD, we analyzed \textit{all} flares from the target period to produce a frequency diagram of the flares by peak irradiance. That is, we created a histogram with logarithmically-spaced bins, normalized by the bin width (Figure \ref{fig:flare_class_hist}). By fitting the result with a power law using the method described in \citet{DHuys2016}, we extract an expression for $dn/dI$, the frequency per peak irradiance, as a function of peak irradiance. 

\begin{figure}
\plotone{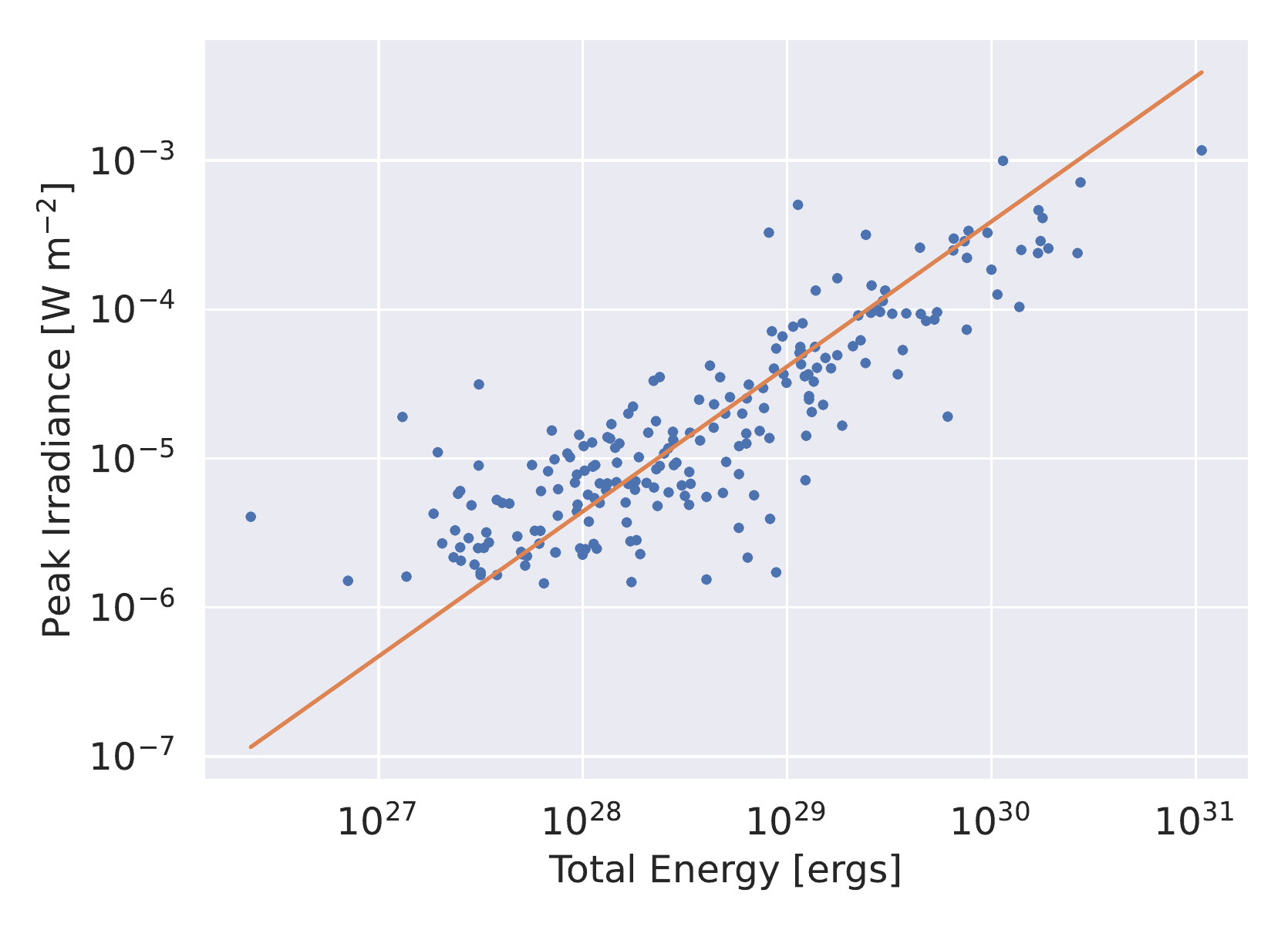} 
\caption{Peak irradiance versus the energy determined with the methodology described in Section \ref{sec:methods}. The slope of the fit shown in red is 0.97$\pm$0.04. These results are comparable to those found in \citet{Veronig2002}.
\label{fig:peak_irradiance_vs_energy}}
\end{figure}

Then, using just the set of flares analyzed by our collaboration, we performed another fit to the peak irradiance as a function of their energies (Figure \ref{fig:peak_irradiance_vs_energy}). From this, we extracted the derivative $dI/dE$, and in turn, computed the desired relationship:

\begin{equation}
\label{eq:unbias}
    \frac{dn}{dE} = \frac{dn}{dI}\frac{dI}{dE}
\end{equation}

Both terms on the right hand side of Equation \ref{eq:unbias} are known -- $\frac{dn}{dI}$ as represented by Figure \ref{fig:flare_class_hist} and $\frac{dI}{dE}$ as represented by Figure \ref{fig:peak_irradiance_vs_energy}. In other words, for every flare we studied, we knew how frequently flares of that peak irradiance \textit{should} occur if there was no bias in our selection method, and could specify that frequency for the particular energy we calculated. Propagation of the standard errors from the two preceding fits produces uncertainties on the resulting $\frac{dn}{dE}$ data points.

To determine $\alpha$, we followed the recommendation made by \citet{DHuys2016}. They note that the MLE method described therein is a special case of a more general Bayesian approach. In order to ingest and propagate the errors on $\frac{dn}{dI}$ and $\frac{dI}{dE}$, we applied that more general Bayesian approach: we used a Gaussian likelihood function with uniform/flat priors to fit the model (Equation \ref{eq:alpha}) to the data. We used a Monte Carlo Markov Chain (MCMC) method with \texttt{emcee} \citep{Foreman-Mackey2013} to sample the posterior distribution of the fit parameters. We show the FFD and line fit in Section \ref{sec:results}. 

\begin{deluxetable*}{cccc}
	\tablecaption{Comparison of GOES/XRS- and MinXSS-derived flare energies}
	\label{tab:spot_check}
	\tablehead{\\
            \colhead{Flare Peak Time (Class)} & \colhead{\begin{tabular}[c]{@{}c@{}}GOES/XRS-derived\\ Flare Energy [erg]\end{tabular}} & \colhead{\begin{tabular}[c]{@{}c@{}}MinXSS-1-derived\\ Flare Energy [erg]\end{tabular}} & \colhead{\begin{tabular}[c]{@{}c@{}}\%\\ Difference\end{tabular}}}
	\startdata
		2016-07-24 06:20 (M2) & $7.47\times10^{28}$ & $6.31\times10^{28}$ & -18\%\\
            2016-11-29 07:10 (C7) & $6.04\times10^{27}$ & $7.81\times10^{27}$ & +29\%\\
            2017-02-22 13:27 (C4) & $1.88\times10^{28}$ & $1.64\times10^{28}$ & -15\%
	\enddata
\end{deluxetable*}

\subsection{MinXSS Analysis Method}
\label{sec:minxss_method}

Of the flares studied in the GOES-15/XRS data, three were also observed by MinXSS-1. Their peak flare times occurred at 2016-07-24 06:20 (M2), 2016-11-29 07:10 (C7), and 2017-02-22 13:27 (C4). We applied the same basic process described above to obtain flare energy for each: 
\begin{enumerate}
    \item Converted from the native units of photons s$^{-1}$ cm$^{-2}$ keV$^{-1}$ to erg s$^{-1}$ cm$^{-2}$ keV$^{-1}$.
    \item Integrated across the 1-8 \AA\ wavelength part of the spectrum in order to match the GOES/XRS long channel bandpass; resulting in erg s$^{-1}$ cm$^{-2}$. 
    \item Identified a pre-flare baseline in the light curve and subtracted it from the light curve.
    \item Identified the start and stop times of the flare (ensuring they were close to those identified in GOES/XRS) and integrated across the corresponding duration; resulting in erg cm$^{-2}$. 
    \item Removed the cm$^{-2}$ by applying 1/r$^2$ for 1 au; resulting in erg. 
\end{enumerate}

Steps 1, 2, and 5 were simple conversions and integrations. Steps 3 and 4 are the key steps requiring some judgement. Because these flares had already been analyzed in the GOES/XRS data, we could simply use the pre-flare time period and flare start/stop times as the guide for the MinXSS flares. 

\section{Results} 
\label{sec:results}

\begin{figure}[h]
\plotone{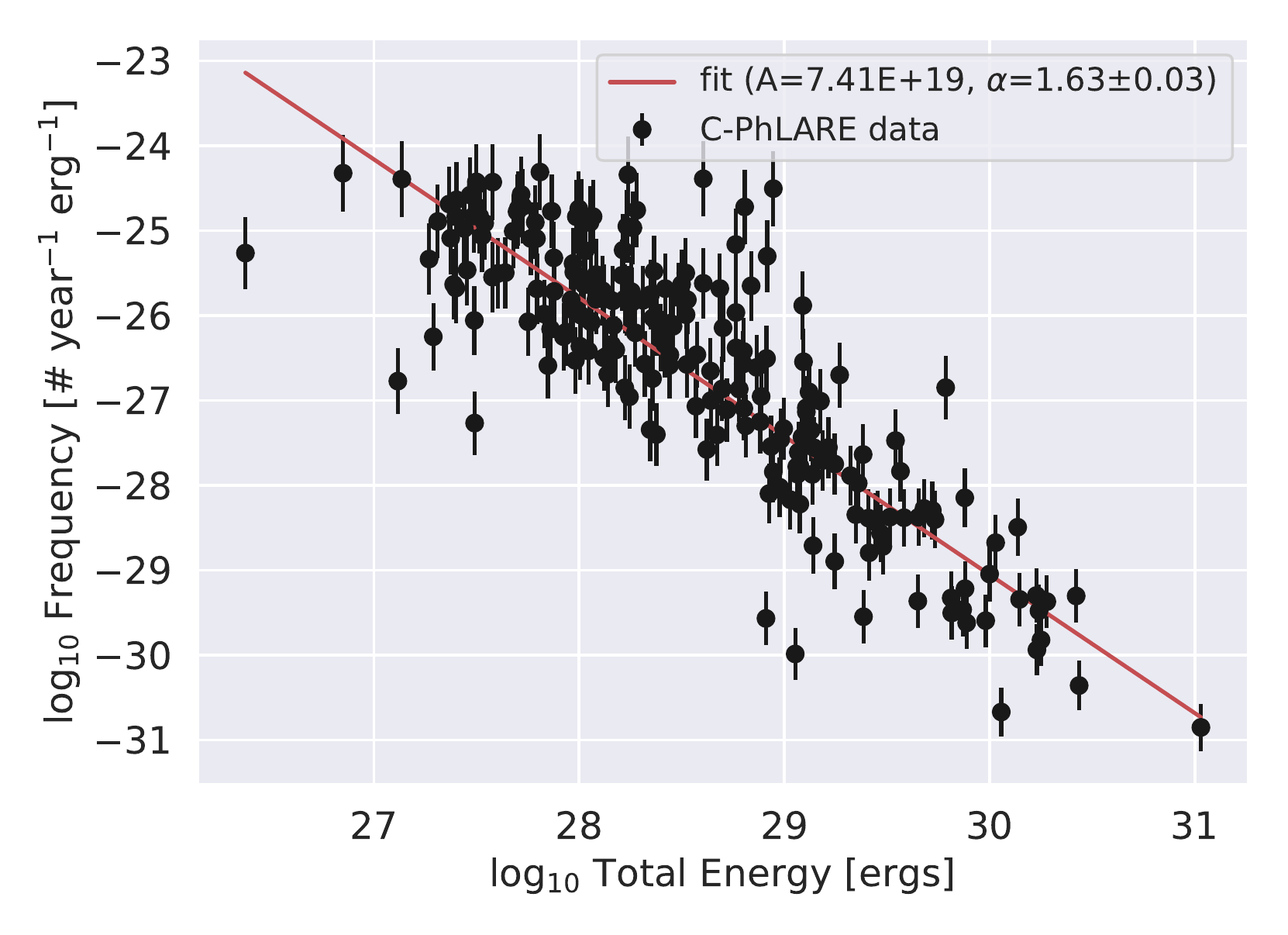} 
\caption{Flare frequency distribution of the C-PhLARE Collaboration, with fit coefficients corresponding to Equation \ref{eq:alpha}. The value of $\alpha$ is below the critical threshold value of 2, suggesting that Alfv\'en waves are an important source of coronal heating.
\label{fig:ffd}}
\end{figure}

The flare frequency distribution, line fit, and slope ($\alpha$) are shown in Figure \ref{fig:ffd}. 

Per Equation \ref{eq:alpha} and the discussion in Section \ref{sec:intro}, the result of $\alpha = 1.63 \pm 0.03$ indicates that even at the outer bound of the uncertainty, $\alpha < 2$, suggesting that Alfv\'en waves, rather than nanoflares, are the dominate mechanism of coronal heating.

\subsection{Cross Checking Flare Energy}
\label{sec:spot_check}

Table \ref{tab:spot_check} shows the comparison of the flare energy as determined with the GOES/XRS data and the MinXSS-1 data. This agreement is within the expected range per \citet{Woods2016a}, accounting for the small irradiance errors introduced when using a flat spectrum in the calibration of the broad-band XRS measurements (see e.g., \citealt{Garcia1994}) and the MinXSS absolute calibration accuracy of about 10\% \citep{Moore2016}. Therefore, we conclude that despite GOES-15/XRS not having an absolute instrument calibration, the resultant flare energies reported in absolute units (ergs) are reasonably accurate. 

\subsection{Comparison With Other Studies} 
\label{sec:comparison}

Table \ref{tab:ffd_compare} places the results from this study in the context of several others. Many solar studies do not agree with each other to within their uncertainties, but this could be explained by differences in methodology as highlighted in Table \ref{tab:ffd_compare}, by uncorrected skews introduced by sub-sample selection effects, or by issues with fitting methods as described by \citet{DHuys2016}. Nevertheless, there appears to be general agreement among solar studies that $\alpha$ is less than the critical value of 2, meaning that these studies conclude that wave coronal heating is dominant. Table \ref{tab:ffd_compare} also shows that, compared to solar studies, many stellar studies have systematically higher values of alpha, even exceeding $\alpha$ of 2 (if only just) and therefore favor the \textit{opposite conclusion}: that nanoflares explain coronal heating. It has been reasonably suggested that the underlying physical mechanism powering solar and stellar flares should be the same (e.g., \citealt{Maehara2017, Airapetian2019}), namely that a convection-driven dynamo generates a magnetic field that is contorted to store energy in the corona and then abruptly freed via magnetic reconnection to produce eruptive events. If this is the case, then the resultant FFD power law should scale well across solar and stellar studies. Table \ref{tab:ffd_compare} shows that this is not the case, but it also makes clear that the methods for all of these studies are not uniform. Moreover, there are other nuances that may need to be accounted for such as the strong dependence of flare frequency on the stellar period of rotation \citep{Notsu2013, Maehara2015, Davenport2016}.

\begin{deluxetable*}{ccccccc}
\tabletypesize{\scriptsize}
\label{tab:ffd_compare}
\tablecaption{Flare frequency distribution comparison}
\tablehead{\\
\colhead{Reference} & \colhead{Target} & \colhead{\begin{tabular}[c]{@{}c@{}}Wavelength\\ regime\end{tabular}} & \colhead{\begin{tabular}[c]{@{}c@{}}Baseline\\ subtraction?\end{tabular}} & \colhead{\begin{tabular}[c]{@{}c@{}}Energy\\ computation?\end{tabular}} & \colhead{Parameter} & \colhead{$\alpha$}
}
\startdata
This study & Sun & SXR & Yes & Yes & Energy & 1.63 $\pm$ 0.03 \\
\citet{Crosby1993} & Sun & HXR & No & Yes & Energy (electrons) & 1.53 $\pm$ 0.02 \\
\citet{Shimizu1995} & Sun & SXR & No & Yes & Energy & 1.5-1.6 \\
\citet{Veronig2002} & Sun & SXR & No & No & Peak irradiance & 2.11 $\pm$ 0.13 \\
\citet{Aschwanden2012} & Sun & SXR & Yes & No & Peak irradiance & 1.98 $\pm$ 0.11 \\
\citet{Caramazza2007} & \begin{tabular}[c]{@{}c@{}}Young stars in \\ Orion Nebula Cluster\end{tabular} & SXR & No & No & Peak irradiance & 2.2 $\pm$ 0.2 \\
\citet{Colombo2007} & \begin{tabular}[c]{@{}c@{}}Cygnus OB2, ONC\\ Nebulae\end{tabular} & SXR & No & No & Peak irradiance & 2.1 $\pm$ 0.1 \\
\citet{Maehara2012} & Sun-like stars & Visible & No & Yes & Energy & 2.0-2.3 \\
\enddata
\tablenotetext{a}{This table is far from comprehensive. For example, \citet{Hudson1991} cites a number of other studies with $\alpha \approx$ 1.8  and \citet{Crosby1993} includes a similar table to this one with $\alpha$ ranging 1.13-2.0.}
\end{deluxetable*}

Table \ref{tab:ffd_compare} also shows how tantalizingly close to $\alpha = 2$ many studies are, some even including the value within their uncertainty. This itself suggests that both processes -- nanoflare heating and wave heating -- are likely important. Beyond the nuances in the data analysis method, the dominance of one process or the other is likely dependent on a variety of factors that are themselves dynamic with time and varying conditions. 

\section{Discussion} 
\label{sec:discussion}

The literature on flare frequency distributions is extensive. The key novelty in the present study is that we had an unprecedented number of data analysts to perform a large number of case studies that we aggregated into a statistical study. In prior studies, it was not feasible to perform pre-flare baseline subtraction and/or energy computation (which requires determining a flare start and stop time) because it is either labor intensive or requires automated algorithms that tend to reject a large fraction of flares. We found a pre-flare subtracted, flare energy frequency distribution with a slope of 1.63 $\pm$ 0.03, which suggests Alfv\'en waves play an important role in heating the solar corona. This result disagrees with much of the prior stellar literature despite the reasonable widespread assertion that the underlying physical processes should be the same for the Sun as other stars. We argue that this discrepancy is likely due to differences in methodology and observational availability. 

We note that, just as in the solar literature, it is difficult to find stellar studies that have performed an energy computation or a baseline subtraction. Solar physicists enjoy a long, nearly unbroken record of observations -- especially in SXR irradiance via GOES/XRS and now also in extreme ultraviolet (EUV) with GOES/EUVS -- but astrophysical studies to date have been primarily limited to time allocation on major facilities, such as \textit{Chandra}/HETGS, which makes the acquisition of complete light curves for flares and pre-flare baselines difficult. An astrophysics mission dedicated to obtaining long baseline observations of stars in the SXR and/or EUV would be highly beneficial to future studies seeking to determine stellar coronal heating mechanisms that could be compared against solar results like those presented here. 

\acknowledgments
Data used for this analysis were processed at the NOAA Space Weather Prediction Center (https://www.swpc.noaa.gov/) and the NOAA National Centers for Environmental Information (NCEI; https://www.ncei.noaa.gov/), and are provided by NCEI at https://www.ngdc.noaa.gov/stp/satellite/goes-r.html. These data were accessed via the University of Colorado's Space Weather Technology, Research, and Education Center's (https://colorado.edu/spaceweather) Space Weather Data Portal (https://lasp.colorado.edu/space-weather-portal).
This work is supported by National Science Foundation Grant Nos. DMR-1548924, PHY-1734006 and NASA Grant NNX17AI71G. 

\vspace{5mm}
\facilities{GOES(XRS), MinXSS}

\software{Code for this paper \citep{CPhlareCode2023},
          AASTeX \citep{AASJournalsTeam2018}, 
	   astropy \citep{Price-Whelan2018}, 
          Google Colab, 
          IDL, 
          IPython \citep{Perez2007}, 
          LyaPy \citep{YoungbloodLyaPy2022},
          matplotlib \citep{Hunter2007}, 
          numpy \citep{Oliphant2006}, 
          pandas \citep{McKinney2010}, 
          R \citep{RLanguage2021},
          scipy \citep{Jonesa2001},
          seaborn \citep{Waskom2021},
          SolarSoft \citep{SolarSoft2012}
          }

\newpage
\bibliography{references}{}
\bibliographystyle{aasjournal}

\allauthors


\end{document}